# Graphene oxide based synaptic memristor device for neuromorphic computing

Dwipak Prasad Sahu, Prabana Jetty and S. Narayana Jammalamadaka*

*Magnetic Materials and Device Physics Laboratory, Department of Physics, Indian Institute of Technology Hyderabad, Hyderabad, India – 502 285.*

*Corresponding author: surya@phy.iith.ac.in

**Abstract**

Brain-inspired neuromorphic computing which consist neurons and synapses, with an ability to perform complex information processing has unfolded a new paradigm of computing to overcome the von Neumann bottleneck. Electronic synaptic memristor devices which can compete with the biological synapses are indeed significant for neuromorphic computing. In this work, we demonstrate our efforts to develop and realize the graphene oxide (GO) based memristor device as a synaptic device, which mimic as a biological synapse. Indeed, this device exhibits the essential synaptic learning behavior including analog memory characteristics, potentiation and depression. Furthermore, spike-timing-dependent-plasticity learning rule is mimicked by engineering the pre- and post-synaptic spikes. In addition, non-volatile properties such as endurance, retentivity, multilevel switching of the device are explored. These results suggest that Ag/GO/FTO memristor device would indeed be a potential candidate for future neuromorphic computing applications.

**Introduction**

The rapid evolution of internet of things for artificial intelligence (AI) and information processing quest a computer hardware which can be operated at low powers and huge volumes of data that needs to be processed with high speed and high efficiency [1,2]. The traditional computing is based on von Neumann architecture, where the computing units are separated with respect to memory units [3]. Present computer hardware has been developed based on complementary metal oxide semi-conductor (CMOS) technology, which faces serious challenges such as implementation of the complicated computation and energy dissipation during its operation. Hence, there is an urgency to advance the present computing methodologies, which needs a change over from compute-centric to data driven processing paradigm [4].

Inspired by the human brain, neuromorphic computing has emerged as an efficient information processing model to address the inherent limitation of von Neumann architecture. In-memory computing, massive parallelism, error tolerant computing and low energy information processing have been some excellent features of neuromorphic computing, which mimic the brain inspired computing [5]. Essentially, the human brain possesses advanced computation ability in an energy efficient way including learning, memory and decision processing [6-7]. The key to simultaneous storage and information processing of the brain is due to the densely connected central nervous system, which consists about $10^{11}$ neurons and are interconnected by about $10^{15}$ synapses. Neurons would be helpful in the information processing and synapses play an important role in learning, memory and self-adaptability of the human brain to complex processes. The connection strength between pre- and post-synaptic neurons, called synaptic weight ($w$), which can be determined based on the amount of neurotransmitter released from neuron to synapse. These synaptic weights can be tuned with time dependent signals and change their strength because of neural action known as 'synaptic plasticity'. This feature has been believed to be the basis for learning and memory processing of the network [3-4]. Two essential neuronal functions for learning process are Long-term potentiation (LTP) and Long-term depression (LTD), where the synaptic weight increases or decreases respectively with respect to time. Spike timing dependent plasticity (STDP) is another important Hebbian learning rule in biological neural network, where the synaptic weight is modulated depending on the relative timing between pre- and post-synaptic neurons.

Recently, RRAM based synaptic devices have garnered much interest owing to it's simple structure, scalability and fast accessing speed. RRAM devices switch between high and low resistance states by applying external electric field and can retain the internal resistance states depending on the history of the bias voltage. Various oxide based RRAM devices such as $HfO_x$, $TiO_x$, $TaO_x$, $NiO_x$, $Pr_{1-x}Ca_xMnO_3$ (PCMO) have been investigated for neuromorphic computing [8]. The mechanism of conductance change in such oxide based synaptic devices has been attributed to the migration of oxygen ions at the interface of active switching layer [9].

Among various 2D materials, graphene oxide is a promising material with excellent downscaling and compatible with conventional silicon-based devices due to atomically thin and weak van der Waals (vdWs) forces between layers [10-11]. Furthermore, the band gap engineering of GO can be modulated by controlling the functional groups on surface, which would allow one to control electrical properties. In recent past, 2D material based RRAM devices have been explored for emulating the synaptic behavior. Romero *et al.* have demonstrated the fabrication of memristor devices based on laser assisted photo-thermal reduction of graphene oxide [12]. In their manuscript, the hysteresis I – V loop has been found to be affected by the laser power. The device showed a stable ON/OFF resistance ratio of ~10 for 10 number of cycles without applying any compliance current. By setting up a compliance value for current, the device could run up to 100 cycles with a ON/OFF resistance ratio of ~ 8. Yet, in another report, the inkjet printed two-terminal GO memory devices have been reported with low (<1) switching voltage with relatively small ON/OFF resistance ratio [13]. The conduction mechanism for the electroforming process has been ascribed to the carbon diffusion in Ag top electrode. While, oxygen ion drift was the main physical mechanism for the subsequent switching cycles in this GO devices. An elaborated switching mechanism of various GO memristive devices has been discussed in [14] based on two models: oxygen ions drift and metal filament formation. A unipolar switching has been reported with silver doped GO memory devices with an endurance of 20 cycles [15]. Shi *et al.* have reported an energy efficient h-BN based synaptic device emulating both LTP and LTD behavior [16]. Essentially, they have studied the dynamic behavior of the synaptic device by applying pulse voltage stress of different amplitude, width and time duration. Feng *et. al,* have demonstrated successfully the learning capability of a mechanically flexible $MoS_2$ synaptic device [17]. Although GO based memory devices show reliable and repeatable switching property, yet much experimental work has not been carried out for brain inspired computing applications. A multi-terminal synaptic device with dual-gate based on $MoS_2$ active layer has been exhibited homosynaptic, heterosynaptic plasticity and other Hebbian learning rules by modulating source−drain voltage and back-gate voltage [18]. Biological synaptic with learning and forgetting experience behavior has been observed in $Ag/Zr_{0.5}Hf_{0.5}O_2$: graphene oxide quantum dots/Ag based memristor device [19]. Yet in another study, Huh *et al.* have explored three-terminal artificial synaptic device based on the $WSe_2$/graphene for neuromorphic computing [20].

Particularly for neuromorphic computing analog switching is important for the continuous and controllable conductance tuning. In addition, analog switching is also important for combining accurate hardware encodings of synaptic weights and data storage. Previous work that has been reported on Au-doped GO devices showed a digital switching where current jumps suddenly from HRS to LRS and thus the device could able to run for 100 pulsed cycles [21]. Sharbati et. al, have demonstrated basic neural functions on electrochemical graphene synapse, where the conductance of the device has modulated by the concentration of Li ions between the layers of the graphene [22]. Although these electrochemical cell devices exhibit low-power synaptic operation yet they used an electrolyte [$LiClO_4$ in poly (ethylene oxide) (PEO)]. In addition, three electrodes (two contact electrodes and one reference electrode) configuration may limit high density integration. Moreover, Li ions can also be a source of contamination for Si devices [23].

Based on the above, in the present manuscript, we have developed GO based RRAM device for the neuromorphic computing applications. Over here we analyzed the synaptic characteristics and also memory capability of the device.

**Experiment:**

Graphene oxide (GO) was prepared by modified Hummer's method, details of which have been published elsewhere [24]. Commercially available fluorine doped tin oxide (FTO) was used as substrate. Prior to device fabrication, FTO glass substrates were cleaned with deionized (DI) water, acetone and 2 – propanol individually for at least 10 min. Initially, a GO solution was prepared by adding 10 mg of GO powder to 10 ml of DI water and ultra-sonicated for 1 hour to obtain a well dispersed solution. The structural characterization of synthesized GO was performed by (Bruker Discover D8) X-ray diffraction (XRD) with Cu Kα radiation ($\lambda = 0.154$ nm). Raman spectroscopy was employed to determine local structure and various functional groups using a Bruker Senterra Laser Micro Raman spectrometer with an excitation source of wavelength 532 nm. Microstructural studies were performed using the field emission scanning electron microscope (Zeiss ultra 55 FE-SEM). Fourier transform infrared spectroscopy (FTIR) was recorded with Bruker Tensor 37 by ATR method in the wavenumber range of 4000 to 500 cm$^{-1}$ with a spectral resolution of 4 cm$^{-1}$ and the absorbance of GO solution was recorded using a Perkin Elmer LAMBDA spectrophotometer between the wavelength range of 200 to 700 nm.

To prepare GO based thin film device, the resulting aqueous solution was drop casted onto the FTO glass substrate followed by annealing at a temperature of 60°C for 30 minutes. Finally, Ag top electrode was deposited using silver conductive epoxy and FTO was used as bottom electrode. The thicknesses of GO film, top Ag contact and bottom FTO electrode were around 1 µm, 0.5 mm and 500 nm, respectively. I – V measurements of the Ag/GO/FTO devices were performed at room temperature using a Keithley 2400 source meter. The synaptic characteristics of the devices were evaluated using a Keithley 4200-SCS semiconductor characterization system and a 4225-PMU ultrafast current–voltage pulse module. During the electrical measurements, positive biased voltage was applied to Ag top electrode and FTO bottom electrode was negatively biased. In the devices, current was sent perpendicular to the plane of the film (CPP) configuration.

**Results & discussion:**

Figure 1(a) illustrates the schematic diagram of a biological synapse which consists of presynaptic neuron, synaptic cleft and postsynaptic neuron respectively. The synaptic characteristics rely on triggering of action potential with release and diffusion of neurotransmitter between pre- and postsynaptic neurons. Figure 1(b) shows the schematic of the Ag/GO/FTO based RRAM device. The morphological characteristics of GO are shown in Figure 1(c), which shows a disordered structure with large amount of wrinkles on the surface. These wrinkles are randomly distributed and formed due to twisting or overlapping of few layers of graphene oxide sheets. The wrinkle shaped sheets are due to the presence of many oxygen-containing functional groups on the surface and edges of GO. The preliminary characterization of the prepared GO is briefed in the supplementary information.

The typical I – V characteristics of the Ag/GO/FTO based electronic synaptic device under DC sweeping mode in the voltage range +3 V to -3 V is shown in Figure 2(a). The device is an electroforming free and the measurements were performed under a current compliance of 1 mA to avoid overshooting of the current. Under the DC sweeping bias of 0 V → +3 V → -3 V → 0 V, the I - V curve shows a hysteresis loop. It is evident that when a positive bias voltage is applied to the top Ag electrode, the device switches from high resistance state (HRS) to low resistance state (LRS), which is termed as SET switching. The device current decreases when a negative bias is applied to top electrode, which returns to HRS, called as RESET switching. Thus the device conductivity changes with positive and negative voltages, this indicates a bipolar resistive switching behavior. It can be observed that the device shows a gradual SET and RESET switching which is essential for neuromorphic computing applications [25].

In order to investigate the conduction mechanism in GO based device, I – V curves were replotted in double logarithmic scale. Figure 2(b) shows the log I – log V plot and its linear fitting of positive biased sweeping region. The HRS consists of three regions namely, Ohmic conduction, trap filled limited behavior and Child's square law [26]. Thus, the current conduction in GO film follows Lampert's theory of space charge limited conduction. At lower bias voltage (< 0.4 V), the slope of the I – V curve is ~ 1.1, which indicates an Ohmic conduction and arises due to more number of thermally generated charge carriers. As the voltage increases, the injected charge carrier density predominant and thus, a trap-filled limited current is observed. When all the traps are filled by the injected charge carriers, the trapped behavior is turned into trap free conduction. This indicates that there is an exponential distribution of charge trapping site in the deeper energy level of GO.

Once all these trapping sites are filled by injected charge carries, it causes an increase in the current and eventually, switches the device to LRS. However, an Ohmic behavior is observed for LRS state as shown in the inset of Figure 2(b). Based on the above mechanism, the RS behavior could be related to formation and dissolution of conductive paths. The microscopic mechanism of change in conductivity in GO film has been explained due to the movement of oxygen induced defects or disorders [27-28]. A schematic illustration of switching mechanism is depicted in the Figure 2(c (i – iii)). The conductivity of GO is associated with amount of high conductivity $sp^2$- bonded carbon atoms and the low conductive $sp^3$- hybridized carbon atoms exists. By the application of electric field, due to chemical reaction or thermal annealing, the $sp^3$ carbon species can be reduced to $sp^2$ species [29-30]. Moreover, the presence of various oxygen-related functional groups such as hydroxyl, epoxide, carboxyl groups in GO induce a deeper energy level distribution of charge traps [27,31]. The absorption and desorption of these functional groups near Ag electrode may be responsible for the change in conductivity of the cell [32-33]. Moreover, first-principle and statistical calculations have also demonstrated the tendency of the oxygen functionalities to agglomerate and form highly oxidized domains surrounded by areas of pristine graphene [34]. Therefore, in the pristine state, the device has higher amount of $sp^3$ domains (reflected as HRS state as shown in Figure. 2 c – (i)) and when a positive bias is applied to the top electrode, $sp^3$ domains creates a large gradients of electrostatic potential. This results in a strong local electric field, which triggers the detachment of functional groups from GO. As a result, more number of $sp^3$- hybridized carbon atoms are reduced to sp$^2$ atoms (Figure 2 c – (ii)) by the electric field. This causes an increase in the conductivity of the device due to higher concentration of interlayer π electron and turns the device into LRS. On the other hand, when a negative voltage is applied to bottom electrode, oxygen ions migrate from the bottom FTO electrode to GO layer.

Generally, FTO electrode serves as an oxygen reservoir and thus, influences the concentration of oxygen vacancies in the oxide layer [35]. Thus, a partial recombination of oxygen ions and vacancies results in the disruption of initially formed conductive path in LRS and turns the device to HRS (Figure 2 – c (iii)). Hong *et al.* have suggested the movement of oxygen in Al/GO/ITO based device as the responsible factor for conduction [36]. However, in another study, dual mechanism of oxygen migration and Al diffusion has been explained for the switching in a similar device, Al/GO/ITO [37]. Therefore, we suggest that the creation and disruption of conductive paths in a highly reduced GO environment may lead to the switching of the device to LRS and HRS respectively.

Figure 3(a) shows the stability of the device on performing 120 continuous switching cycles. It can be inferred that the device exhibits nearly stable switching which can be useful to emulate the synaptic characteristics similar to the biological synapse. The electrical characterization of memristive devices depends on the voltage sweep rate as an increased voltage ramp rate, which results in faster movement of electrons through the oxygen vacancies of the 1D conductive filaments. The sweep rate for the I - V curve in Figure 3(a) is 0.05 V/s. To further examine the stability of the switching performance, the endurance test of the device was performed and the results are shown in the inset of Figure 3(a). It can be seen that the device shows a good ON/OFF resistance ratio ($R_{OFF}/R_{ON}$ = 22) even after 100 DC sweeping cycles. Further, the endurance performance is enhanced by applying a pulsed voltage. As shown in Figure 3(b), a pulsed voltage of +1.5 V/-1 V with a pulse width of 100 µs was applied to the device. A higher number of switching cycles (~450 cycles) with an increased memory window of $R_{OFF}/R_{ON}$ = 34 is observed under a pulsed bias. The reason behind degradation of the device above 450 cycles could be attributed to the enhanced Joule heating effect after continuous switching cycles. This results in permanent filament rupture leading to device failure. Our endurance cycles are much better (~ 450 cycles) if we compare with other GO based devices. For example, in Pt/GO/ITO memristor, the device can well differentiate between LRS and HRS for 100 cycles with a wide variation in HRS values [38]. A similar number of switching cycles have been observed in Al/GO/Al film deposited on flexible polyethersulfone (PES) substrate [39]. A stable ON/OFF ratio has been demonstrated in Al/GO/ITO devices with 100 switching cycles [40].

The power consumption of the devices for SET switching ($P_{SET}$) is ~0.045 mW and RESET switching ($P_{RESET}$) is ~2.5 mW. A higher value of power consumption has also been reported for Sn/HfO$_2$/Pt device with $P_{SET}$ = 3.5 mW and $P_{RESET}$ = 10 mW respectively[41]. In NiO thin film based unipolar devices (Nb/NiO/Nb), the SET power (12.3 mW) and RESET power (5.7 mW) have been reported under a compliance current of 15 mA [42]. Yet on another device, Ta/ Ta$_2$O$_5$/Pt the power consumed has been reported as 2.3 mW and 8 mW for SET and REST switching respectively [43].

We also investigated the stability of resistance both in HRS and LRS with time. Figure 3(c) indicates the retention characteristics of the resistance states over a time period of $10^4$ sec. A multilevel switching event is shown in the Figure 3(d) by varying the compliance current. It can be observed that the resistance in the LRS decreases as the compliance current increases. This could be due to formation of more stable and strong conductive filaments (CF), which cannot be ruptured during RESET switching [44]. Thus, CF sizes can be effectively controlled by the compliance current [25].

In order to explore an RRAM device as a synaptic device, pulse programming is an effective way to mimic the synaptic behavior and a reliable method for practical application in neuromorphic computing. Therefore, we have demonstrated the synaptic learning such as synaptic plasticity caused by SET and RESET switching in the present GO based device through long term potentiation (LTP or gradual increase in conductance) and long term depression (LTD or gradual decrease in conductance). Figure 4 demonstrates the gradual conductance modulation by applying identical positive or negative pulses. The shape of applied pulse bias and the resulting electrical output is illustrated in Figure 4(a) and 4(c). The amplitude was fixed at 1.2 V for LTP and -1.2 V for LTD with a pulse width of 1 ms. It can be observed that there is a continuous increase in the conductance for positive pulse voltage which mimics LTP [Figure 4(b)] and a decrease in the conductance for negative pulse, which indicates the LTD [Figure 4(d)].

In addition, the potentiation and depression is also demonstrated at three different pulses 1 V, 1.2 V and 1.4 V with a pulse width of 10 µs. As shown in Figure 5(a), the device shows non-linear characteristics of conductance change with different pulse amplitude. The potentiation characteristics for pulse heights of 1 V and 1.2 V show a moderate change in the initial cycles and afterwards it increased gradually. However, with the pulse amplitude of 1.4 V, initially the current increases rapidly and reaches the maximum conductance up to 25 cycles and afterward it decreases marginally. The decrease in conductance could be attributed to the Joule heating effect which causes disruption in gradual filament formation due to high temperature [45]. The depression behavior is also achieved by setting different pulse amplitudes as shown in the Figure 5(b). Therefore, the conductance modulation of GO based device to achieve LTP and LTD characteristics gives an indication that GO based RRAM device may be a probable candidate for the neuromorphic computing applications. It has been reported that in case of GO, due to low density of states (DOS) near the Fermi level, the conductivity is less in comparison with the reduced graphene oxide [46]. The low conductivity in our present device could be due to the low DOS. On top of that leakage current may further reduce the conductance in LTD measurements.

In terms of the dynamical evolution under applied electric fields, we believe that electric field would play a major role in controlling the DOS at the Fermi level. However, it would be very difficult to say how much shift exists for Fermi level with the applied electric field as complicated things are involved such as leakage currents, oxygen ion migration and joule heating effect *etc*. In summary, there is a correlation between conductance, electric field and band structure of graphene oxide due to which we do see a significant change in transport properties. This change in conductance also involves other effects as we stated above.

On top of that if we observe the conductance values in LTP and LTD, they are not same, which can be ascribed to leakage currents. The leakage current in RRAM devices has been ascribed to oxygen migration [47-49]. The leakage current increases as the concentration of electrons enhance in the switching layer due to removal of oxygen ions under an applied bias voltage. The back diffusion of oxygen ions into the switching layer under reverse bias voltage reduces the leakage current in the device. Thus, in our devices, as the magnitude of reverse bias voltage increases, the electron injection into the GO layer decreases and thus may reduce the leakage current.

Generally, a gradual SET and RESET switching (analog switching) is more preferable than that of an abrupt conductance change (digital switching). This is because multilevel intermediate states cannot be effectively achieved by abrupt switching [25]. The gradual SET process in analog switching is believed due to formation of larger diameter filaments or multiple filaments with a higher current in the switching material. On the other hand, partial annihilation of few CFs or dissolution of a larger region of the filament leads to a gradual RESET switching [50]. To emulate the non-linear transmission characteristics of synapse and further to demonstrate the analog behavior of the device, cyclic RS study was carried out by varying the RESET voltage ($V_{RESET}$) from -0.8 V to -1.5 V with a step of 0.1 V while maintaining a compliance current of 1 mA. As shown in the Figure 5(c), the device gradually switched to higher resistance states as the $V_{RESET}$ increases. This behavior imitates the "forgetting" process and analogous to the depression behavior of biological synapse. Thus, as the RESET voltage decreases, the resistance of HRS increases, which hints that intermediate resistance state can be achieved by controlling the voltage. Similarly, various conductance states can be achieved by applying consecutive SET voltage from 0.8 V to 1.5 V with a step voltage of 0.1 V and with a current compliance of 5 mA as shown in the Figure 5(d). Indeed, this behavior mimics the learning process or potentiation characteristics in neuromorphic computing system.

Furthermore, we have investigated the spike-timing-dependent-plasticity (STDP) rule which is another important learning rule in the neural system. STDP refers to the conductance modulation depending on the relative timing of pre-synaptic spike and post-synaptic spike. If the pre-spike precedes the post-spike, then the synaptic weight increases and the synapse undergoes a long term potentiation. On the contrary, if the post-spike precedes the pre-spike, synaptic weight decreases leading to long term depression. The change in conductance is calculated using the formula $\Delta G = (G_{after} − G_{before})/ G_{before}$, where $G_{before}$ is conductance before pre- and post-spike pair and $G_{after}$ is conductance after pre- and post-spike pair. The measurement is performed by a pre-synaptic pulse to the top electrode and an identical post-synaptic pulse to the bottom electrode. Figure 6(a) shows the schematic of pulse that is used for STDP experiment. The initial conductance value, $G_{before}$, for LTP event is 581 μS and for LTD event is 554 μS. The applied pre- and post-pulse signals of amplitude 1 V with a time delay of 5 μs is shown in Figure 6(b). The STDP characteristics of synaptic device is depicted in Figure 6(c). It is evident that the conductance increases as the relative timing between two spikes is short. Based on the above results, it can be concluded that Ag/GO/FTO based device shows STDP learning behavior with a change in conductance depending on the relative time similar to the biological synapse.

**Summary:**

In summary, we have investigated the RS phenomena and synaptic behavior of the Ag/GO/FTO based memristor device. The device showed stable analog switching for 120 DC sweeping cycles with a good ON/OFF resistance ratio and further improvement in switching is demonstrated by applying pulsed voltage. For synaptic behavior study, multistate storage is achieved by varying compliance current, RESET and SET voltage. The long term potentiation (LTP) and long term depression (LTD) behavior is achieved by applying a constant pulse amplitude. The conductance modulation is achieved by varying the pulse height. STDP characteristics are demonstrated by applying a pair pulses with a time delay between them.

We would like to acknowledge Indian Institute of Technology, Hyderabad for providing financial support. The author Dwipak Prasad Sahu is thankful to the Department of Science and Technology, India (DST-INSPIRE) for the award of senior research fellowship (SRF).

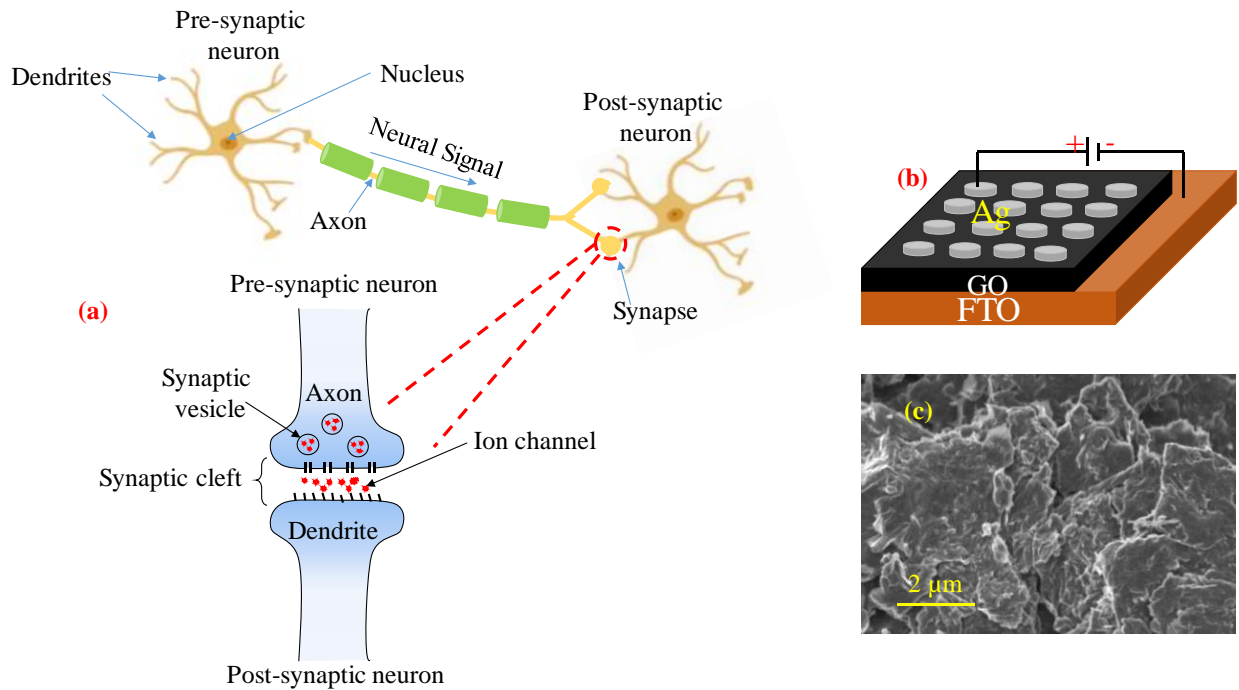

**Figure 1: (a)** A schematic diagram of a biological synapse **(b)** Schematic illustration of Ag/Graphene oxide (GO)/FTO based RRAM device. Here, Ag top electrode is positively biased and FTO is negatively biased in all the electrical measurements. **(c)** Morphological structure of GO indicates a wrinkle shaped sheets.

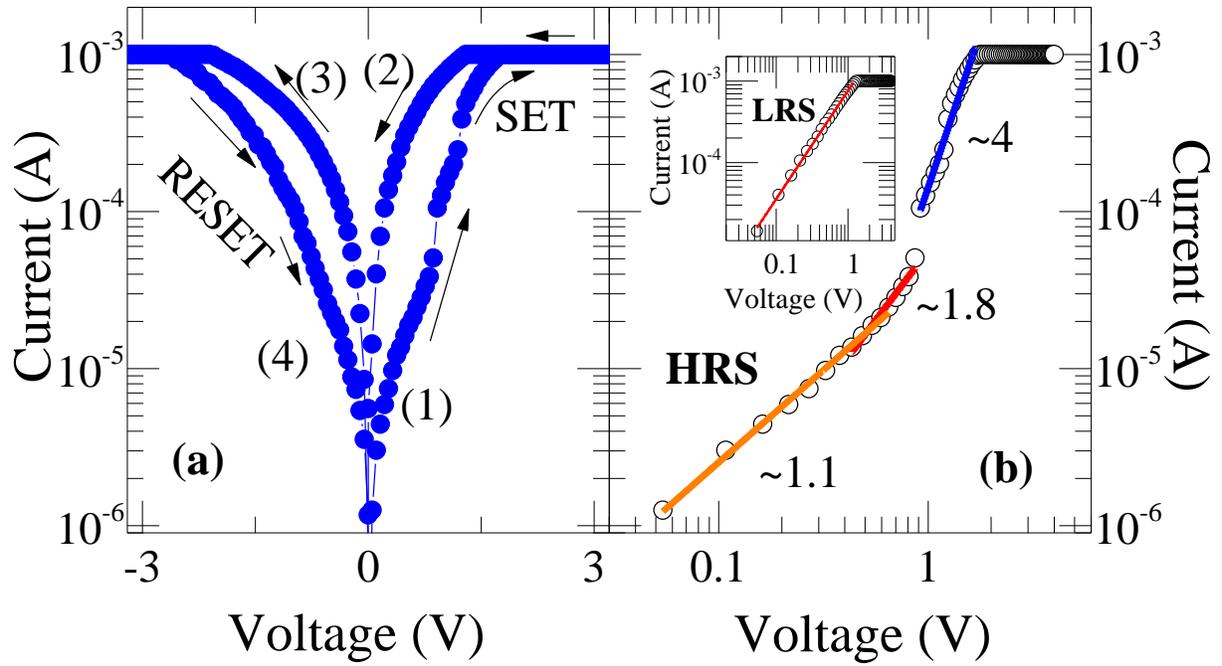

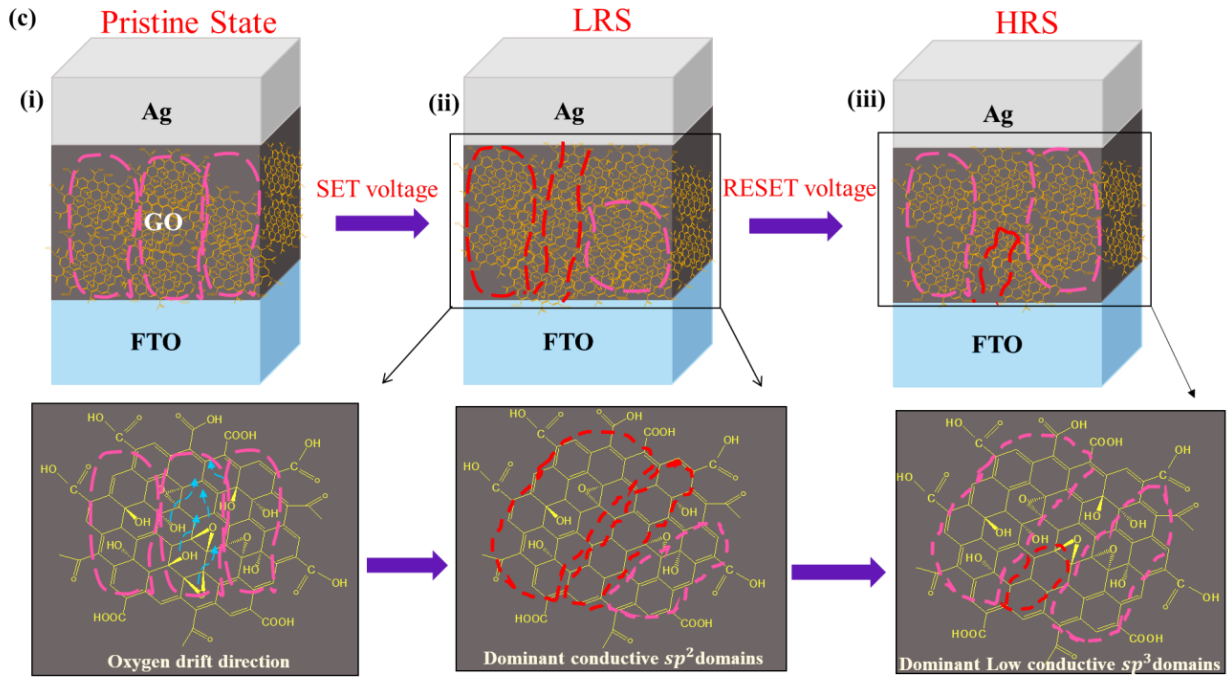

**Figure 2: (a)** I – V characteristics of Ag/GO/FTO synaptic device measured under a DC sweeping with voltage sweep of +3 V to -3 V and with 1 mA compliance current **(b)** double logarithmic plot of I – V characteristics pertinent to HRS which indicates trap filled space charge limited

conduction and inset shows Ohmic nature of LRS **(c)** Schematic illustration of proposed switching mechanism in Ag/GO/FTO device. Initially, the device is at high resistance state due to $sp^3$ domains (pink color) as shown in **c – (i)**. Under the effect of positive bias voltage, a large gradient of electrostatic potential is applied to the non-conductive regions which results in a strong local electric field. This could induce oxygen ion drift (blue color) and reduces $sp^3$ atoms (pink color) to form dominant $sp^2$ domains (red color) as shown in **c – (ii)**. As a result, a percolation path is formed, which leads to LRS. In the reverse bias, a partial recombination of oxygen ions and vacancies results in the disruption of initially formed conductive path in LRS and turns the device to HRS (due to dominant $sp^3$ atoms as shown in **c – (iii)**).

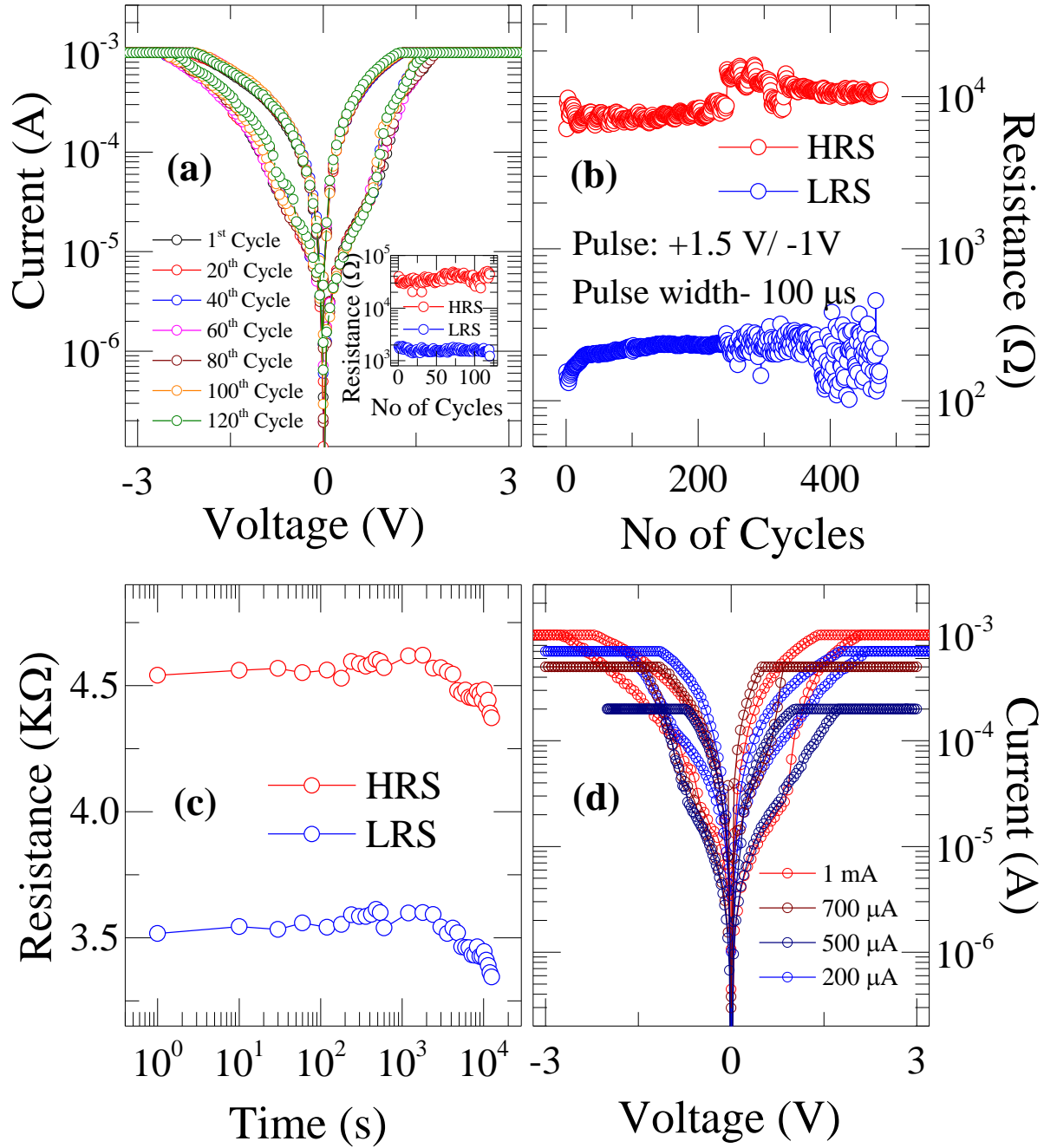

**Figure 3:** **(a)** I – V characteristics of the device for 120 cycles. Inset shows endurance performance of Ag/GO/FTO device for 120 cycles by DC sweeping mode **(b)** Endurance characteristics with ~450 cycles by applying pulse voltage of +1.5 V/-1 V and with a pulse width of 100 µs **(c)** Retention characteristics over a time period of $10^4$ sec. **(d)** Multilevel state switching under different compliance currents.

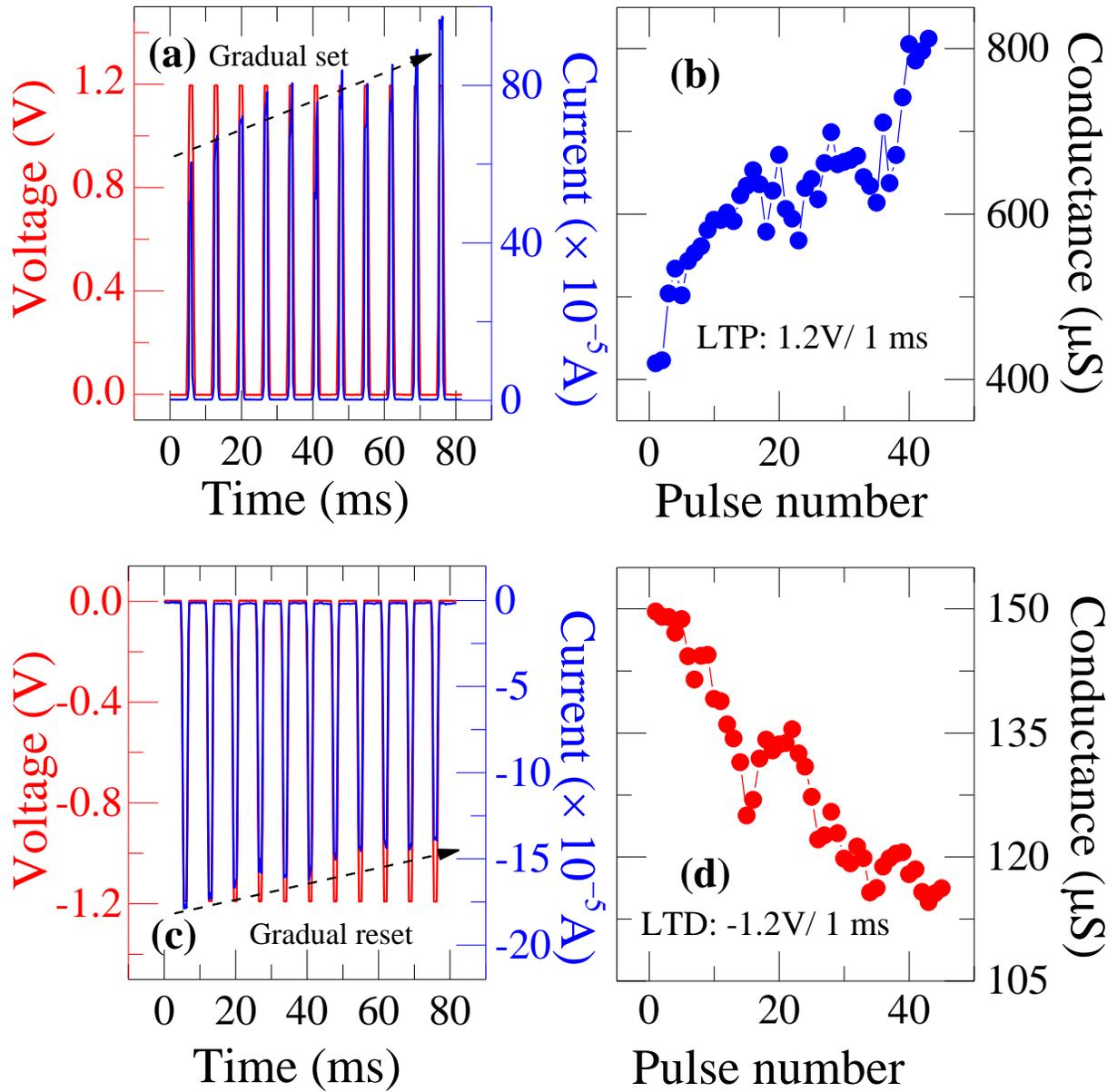

**Figure 4: (a)** Conductance modulation by applying a constant positive pulse amplitude of 1.2 V with a pulse width of 1 ms. **(b)** gradual SET process indicating long term potentiation (LTP) characteristics **(c)** conductance modulation by applying a negative pulse (-1.2 V with a 1 ms pulse width) **(d)** gradual RESET process, which indicates long term depression (LTD) characteristics.

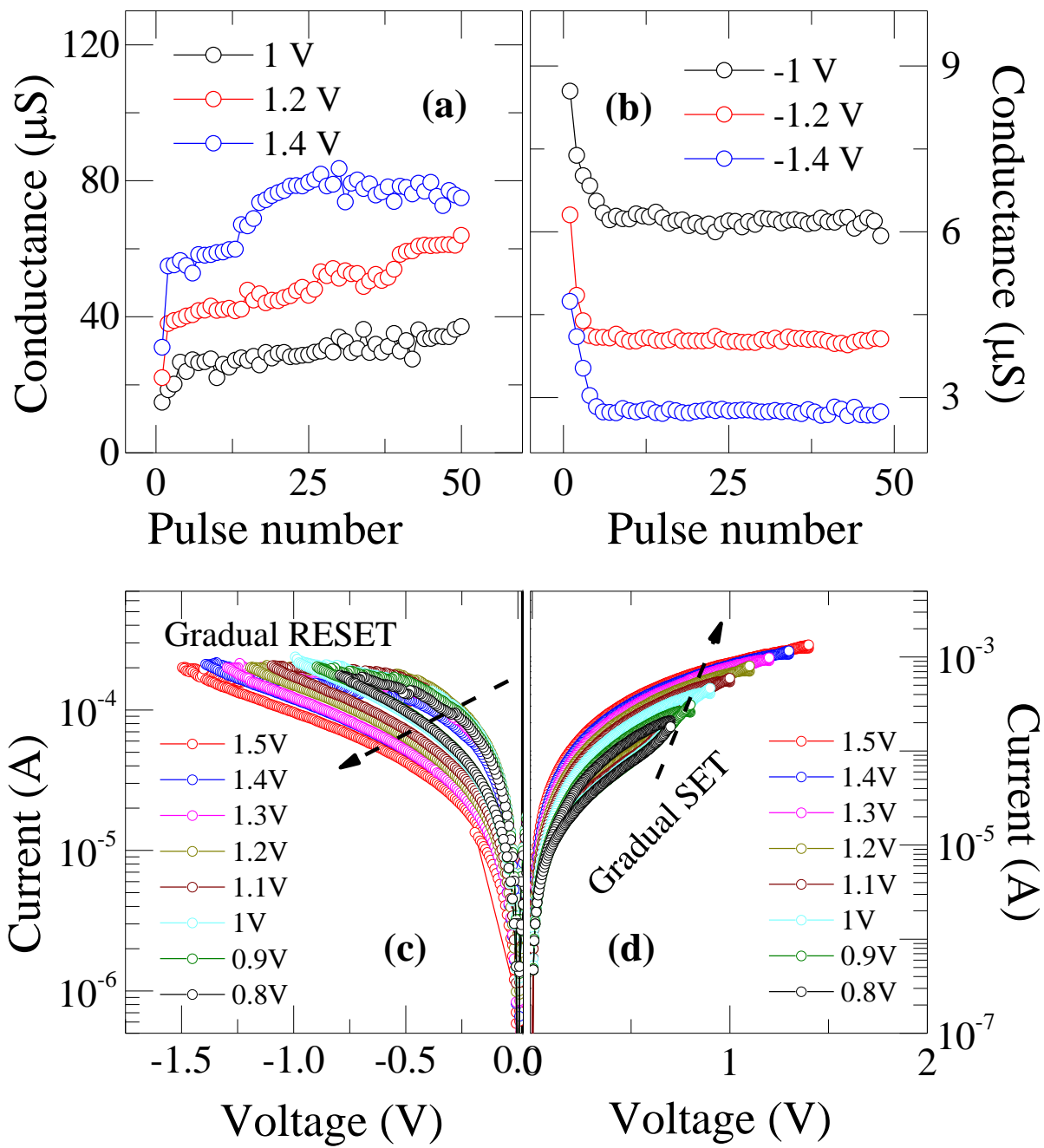

**Figure 5:** conductance change of Ag/GO/FTO device during **(a)** potentiation and **(b)** depression for different pulse amplitudes measured for 50 consecutive pulse trains. **(c)** gradual RESET process by varying $V_{RESET}$, which indicates the depression characteristics **(d)** potentiation behavior of the synaptic device by varying $V_{SET}$.

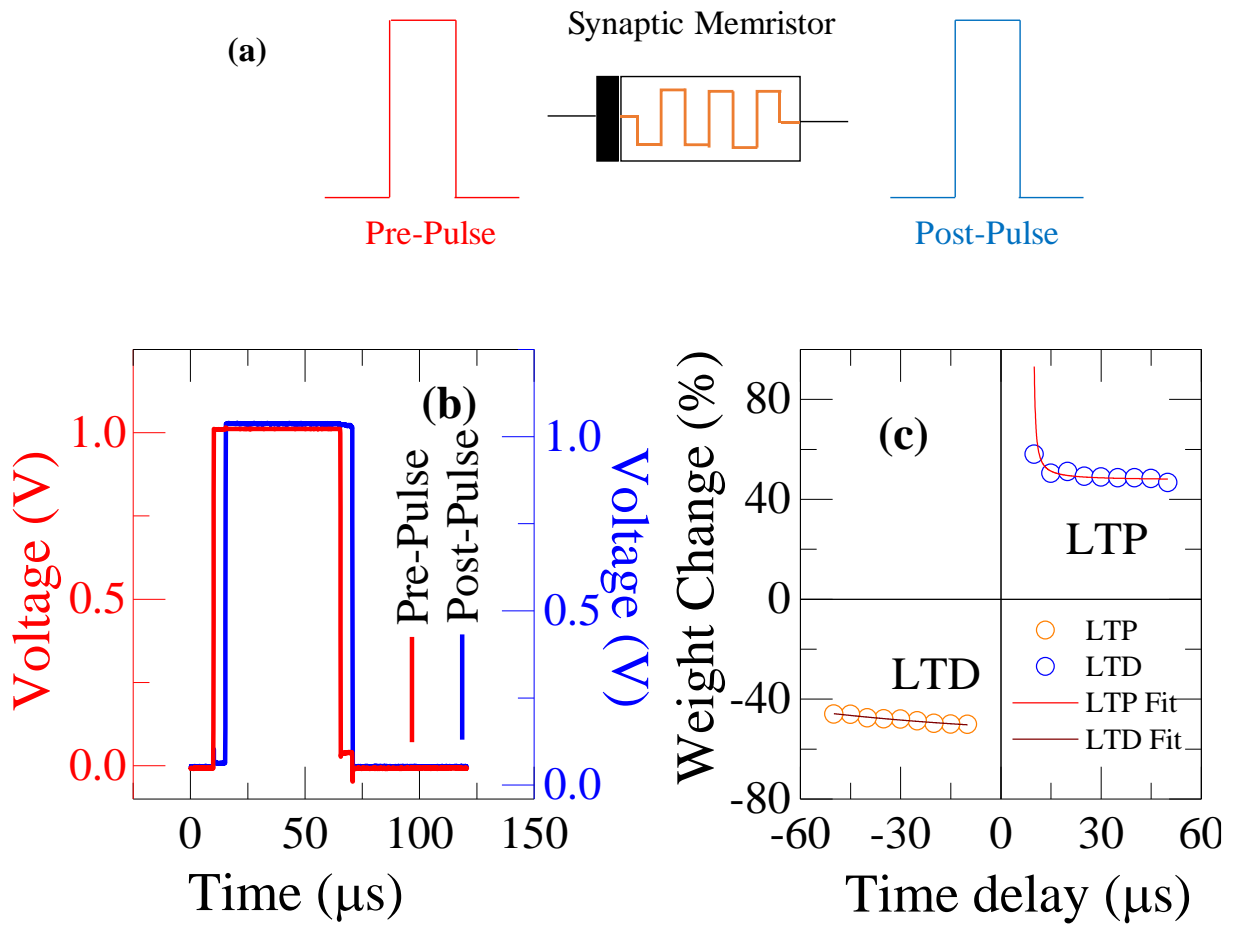

**Figure 6: (a)** Schematic presentation of STDP set up and pulse used for experiment **(b)** the applied pre- and post-pulse signals with a pulse amplitude of 1 V and time delay of 5 µs **(c)** STDP characteristic of the device which shows the relationship between time delay with synaptic weight change.